\documentclass[aps,pra,twocolumn,showpacs,superscriptaddress,nofootinbib]{revtex4-1}

\usepackage[utf8x]{inputenc}
\usepackage{ucs}
\usepackage{amsmath}
\usepackage{amsfonts}
\usepackage{amssymb}
\usepackage{makeidx}
\usepackage{cellspace,booktabs}
\usepackage{natbib}
\usepackage{lipsum}
\usepackage{bm}

\usepackage[usenames, dvipsnames]{color}
\definecolor{light-gray}{gray}{0.55}

\usepackage{microtype}

\usepackage{graphicx}

\usepackage{hyperref}

\renewcommand{\dag}{^{\dagger}}
\newcommand{\exv}[1]{ \langle #1 \rangle }

\newcommand{\bra}[1]{ \langle #1 \rvert }
\newcommand{\ket}[1]{ \lvert #1 \rangle}

\newcommand{\wa}{\omega_a}
\newcommand{\wc}{\omega_r}
\newcommand{\wdr}{\omega_d}

\newcommand{\pfrac}[2]{\frac{\partial #1}{\partial #2}}

\newcommand{\nRK}{n_{\mathrm{RK}}}

\begin{document}

\date{\today}
\author{Samuel Boutin}
\affiliation{Institut Quantique and D{\'e}partement de Physique, Universit{\'e} de Sherbrooke, Sherbrooke, Qu{\'e}bec, Canada J1K 2R1}
\author{Christian Kraglund Andersen}
\email{E-mail: chanders@phys.ethz.ch}
\affiliation{Department of Physics and Astronomy, Aarhus University, DK-8000 Aarhus C, Denmark}
\affiliation{Department of Physics, ETH Zurich, CH-8093 Zurich, Switzerland}
\affiliation{Institut Quantique and D{\'e}partement de Physique, Universit{\'e} de Sherbrooke, Sherbrooke, Qu{\'e}bec, Canada J1K 2R1}
\author{Jayameenakshi Venkatraman}
\affiliation{Department of Physics, Indian Institute of Technology Kanpur, Kanpur, 208016, India}
\author{Andrew J. Ferris}
\affiliation{ICFO-Institut de Ciencies Fotoniques, The Barcelona Institute of Science and Technology, 08860 Castelldefels (Barcelona), Spain}
\author{Alexandre Blais}
\affiliation{Institut Quantique and D{\'e}partement de Physique, Universit{\'e} de Sherbrooke, Sherbrooke, Qu{\'e}bec, Canada J1K 2R1}
\affiliation{Canadian Institute for Advanced Research, Toronto, Canada}

\title{Resonator reset in circuit QED by optimal control for large open quantum systems}

\begin{abstract}
We study an implementation of the open GRAPE (Gradient Ascent Pulse Engineering) algorithm well suited for large open quantum systems. While typical implementations of optimal control algorithms for open quantum systems rely on explicit matrix exponential calculations, our implementation avoids these operations leading to a polynomial speed-up of the open GRAPE algorithm in cases of interest. This speed-up, as well as the reduced memory requirements of our implementation, are illustrated by comparison to a standard implementation of open GRAPE. As a practical example, we apply this open-system optimization method to active reset of a readout resonator in circuit QED. In this problem, the shape of a microwave pulse is optimized such as to empty the cavity from measurement photons as fast as possible. Using our open GRAPE implementation, we obtain pulse shapes leading to a reset time over four times faster than passive reset. 
\end{abstract}

\maketitle

\section{Introduction}
Optimal control, which aims at devising ideal control pulses to optimize a given physical process, is finding wide applications in the fields of theoretical quantum information science~\cite{PhysRevLett.89.188301, PhysRevA.74.022312, schirmer09, PhysRevA.84.042315, PhysRevLett.103.110501, 0953-2048-27-1-014001, PhysRevA.90.052331, PhysRevLett.106.190501, PhysRevA.84.022307}, quantum optics~\cite{glaser2015training} and quantum chemistry~\cite{shapiro2003principles} amongst other quantum fields~\cite{brif10review}. Quantum optimal control theory has also found applications in the laboratory, in particular with nuclear magnetic resonance~\cite{nmrReview2007}, trapped ions~\cite{PhysRevA.77.052334} and superconducting qubits~\cite{PhysRevA.82.040305, PhysRevLett.112.240504}. In most instances, optimal control is applied to unitary processes where dissipation is a nuisance and is considered to be detrimental to the desired process. If properly engineered, dissipation can, however, be a useful resource for tasks ranging from quantum state preparation in circuit QED~\cite{didier:2014a, liu:2016a} to universal quantum computation~\cite{verstraete:2009a}. While not as widespread as its dissipation-less version, open quantum optimal control has also been studied~\cite{PhysRevA.78.012358, PhysRevA.84.022305, schulte11opengrape, floether12open, goerz14open, koch2016controlling}, with the most widely used algorithms being the open system versions of the GRAPE (Gradient Ascent Pulse Engineering)~\cite{Khaneja2005296,schulte11opengrape} and Krotov~\cite{krotov1995global, yvon03krotov} algorithms, while other optimization algorithms \cite{PhysRevA.84.022326, PhysRevA.92.062343, engel2009local} may also prove useful in the context of open systems.

An important difficulty when dealing with open quantum systems is that the Schr\"odinger equation is replaced by a master equation and the wavefunction by a density matrix~\cite{gardiner:2004b}. For a system of dimension $d$, described by the master equation $\dot\rho = \hat{\mathcal{L}} \rho$, a standard approach for optimal control is then to express the density matrix $\rho$ as a vector $\rho_L$ of dimension $d^2\times1$ and the superoperator $\hat{\mathcal{L}} \cdot$ representing the master equation as a matrix $L$ of size $d^2 \times d^2$~\cite{schulte11opengrape}. In this representation, time evolution can be obtained by direct matrix exponentiation which, given the large size of $L$ even for moderate $d$, rapidly becomes numerically intensive. While alternative implementations with optimized time propagators, for example using expansion in Newton polynomials~\cite{goerz2015optimizing,ashkenazi95newton} or by projection onto Krylov subspace~\cite{gutknecht2007brief, tal2007restart} can be used, they lack the simplicity of the direct matrix exponentials and are thus not as widespread. Optimal control in open quantum systems has therefore been mostly limited to systems with small Hilbert space size. Here, we present an alternative implementation of the open GRAPE algorithm that eliminates the need to generate the large matrix $L$. This implementation is well suited for large open quantum systems and avoids explicit matrix exponentiation by rather relying on simple and standard Runge-Kutta time-integration of the master equation.

As an example, we apply this open GRAPE implementation to a problem of current experimental interest: resonator reset in circuit QED. In this architecture, qubit readout is realized by injecting microwave photons in a resonator, which is dipole coupled to qubits, and by measuring the photons reflected or transmitted by the resonator. With excess photons in readout resonators having been shown to be a source of unwanted coherent \cite{PhysRevA.93.012316, PhysRevApplied.4.054001, PhysRevA.94.012347} and incoherent \cite{PhysRevA.77.060305, PhysRevB.86.180504, PhysRevB.86.100506} qubit transitions, it is essential to reset the system by removing the measurement photons from the resonator after readout, and before further coherent manipulations or subsequent readout of the qubit can be performed. The usual approach is to wait for several photon decay times $T_\kappa = 1/\kappa$, with $\kappa$ the resonator decay rate, for the photons to leak out of the resonator~\cite{mcclure2015rapid, bultink:2016a}. In practice, this is, however, often too slow as a fast repetition time of qubit measurements is critical, e.g., for quantum error correction~\cite{PhysRevA.86.032324}. With this standard passive approach, this need for fast decay is in contradiction with the necessity to use high-Q resonators to avoid qubit Purcell decay~\cite{houck:2008a}. Alternatively, active reset can be performed, where a microwave tone is used to empty the resonator in a shorter time. Such a reset tone can be either conditional on the readout result~\cite{bultink:2016a} or unconditional~\cite{mcclure2015rapid, bultink:2016a} using no knowledge of the resonator and qubit states. Devising an active unconditional reset protocol is an ideal test problem for our open GRAPE implementation since it is an intrinsically dissipative process requiring a large Hilbert space size due to the many resonator photons used for qubit measurement. 
Moreover, active resonator reset in circuit QED was recently explored experimentally~\cite{mcclure2015rapid, bultink:2016a}, giving us the opportunity to consider parameters of current practical interest. 
In addition to this example, the numerical approach presented here has also recently been successfully applied by some of us to optimize a fast initialization of cat states in a Kerr resonator based on two-photon driving~\cite{Puri:2017ee}.

The paper is organized as follows: We first present a brief overview of open GRAPE in Sec.~\ref{sec:grape}. We then discuss our implementation of this algorithm in Sec.~\ref{sec:implementation}. Section~\ref{sec:reset} is devoted to the application of the algorithm to active resonator reset. Finally, Sec.~\ref{sec:conclusion} summarizes our work. 

\section{Optimal control for open quantum systems}
\label{sec:grape}

Before discussing our implementation of the algorithm, we first present an overview of the problem solved by the GRAPE algorithm~\cite{Khaneja2005296} and of open GRAPE~\cite{schulte11opengrape}. The reader familiar with these concepts can immediately skip to Sec.~\ref{sec:implementation}.

\subsection{The control problem}

Consider a system with the free Hamiltonian $H_0$ and subject to $R$ independent control fields each described by the Hamiltonians $H_k$ such that the full system Hamiltonian reads~\cite{PhysRevA.84.022307,Khaneja2005296}
\begin{equation}
	H(t) = H_0 + \sum_{k=1}^R u_{k}(t) H_k.
	\label{eq:Hamiltonian}
\end{equation}
The classical parameters $u_k(t)$ in the above expression can be continuously adjusted to change the strength of the control fields on the system. In the context of circuit QED, these $u_k(t)$ can, for example, correspond to the time-dependent amplitude of different microwave drives on the resonator or the qubit.

\begin{figure}[t]
\includegraphics[width=0.95\linewidth]{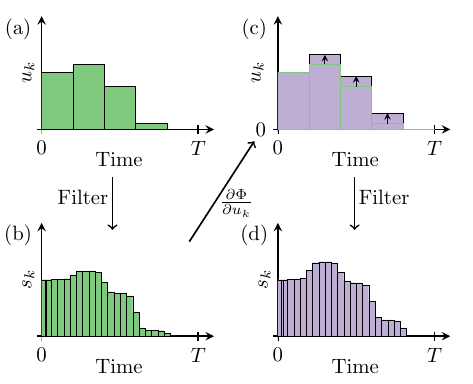}
\caption{Schematic of a gradient-based optimization step with GRAPE-type controls update, and Gaussian filtering to account for experimental constraints.	Starting from initial controls shown in (a), we calculate the filtered experimental pulse shape in (b). From this filtered field, the gradient, $\partial \Phi / \partial u_k$, is calculated using the chain rule (see Ref.~\cite{PhysRevA.84.022307} or App.~\ref{sec:gaussian} for details) and the controls are updated in (c), which leads to a new filtered field in (d).	The boundary conditions of the field are taken into account by fixing the first and last control.} \label{fig:filtering}
\end{figure}

The objective of the control problem is to find the optimal set $\left\{ u_k(t)\right\}$ to accomplish a specific task, most typically implementing quantum gates~\cite{PhysRevLett.103.110501, 0953-2048-27-1-014001}. This can be expressed as an optimization problem where the goal is to maximize the performance index $\Phi[\lbrace u_k \rbrace]$, a measure for the success of the desired task and a functional of the control parameters. As the optimization problem must be of finite dimension, the control amplitudes, $u_k(t)$, are taken to be piecewise constant. For a process of duration $T$, each $u_k(t)$ is divided into $N$ time steps of duration $\Delta t = T/N$ as illustrated in Fig.~\ref{fig:filtering}(a). In this way, for the $j^{\mathrm{th}}$ step, i.e. for \mbox{$t \in [(j-1)\Delta t; \; j\Delta t )\,$}, the function $u_{k}(t)$ is a constant of amplitude $u_k(j)$ with $j \in \lbrace 1,2, \ldots\, , N\rbrace$. The elements of the set $\left\{ u_k(j) \right\}$ are referred to as the controls.

In practice, these sharp controls are smoothed out by the finite bandwidth of the control lines. Following Ref.~\cite{PhysRevA.84.022307} and as illustrated in Fig.~\ref{fig:filtering}(b), this important experimental consideration can be taken into account by filtering the controls in the evaluation of the performance index and its gradient. This filtering procedure maps the piecewise constant functions described by the set $\left\{ u_k(j) \right\}$
to smoother piecewise constant functions defined by the larger set $\left\{ s_k(l) \right\}$
with $l= 1,2, \dots M$ and $M  = T / \delta t\gg N$. For completeness, details of this filtering procedure can be found in Appendix~\ref{sec:gaussian}.

An approach to optimize the performance index is to update the controls by using a gradient-based optimization algorithm such that~\cite{Khaneja2005296}
\begin{align}
u_k(j) \rightarrow u_k(j) + \sum_{lm}B_{kj,lm} \frac{\partial \Phi }{\partial u_l(m)}, \label{eq:update}
\end{align}
where $B_{kj,lm}$ are the elements of a step matrix which depends on the details of the chosen optimization algorithm. Simple gradient descent optimization corresponds to the choice $B_{kj,lm} \propto \delta_{kl}\delta_{jm}$, while for more sophisticated methods, such as the  Broyden-Fletcher-Goldfarb-Shanno (BFGS) algorithm, $B_{kj,lm}$ is related to the inverse of the Hessian matrix~\cite{Fletcher:2000uq}. Since the BFGS algorithm leads to improved convergence~\cite{Fouquieres:2011fk}, it will be used in the numerical computations presented below. A non-trivial step in the update rule Eq.~\eqref{eq:update} is the evaluation of the gradient of the performance index. While this can be done by numerical derivatives, this approach become intractable for problems with a large set of controls. 
Using an analytical result described below for open systems, the GRAPE algorithm allows for an efficient calculation of this gradient.

\subsection{Open GRAPE}

We consider an open quantum system whose dynamics is described by the Markovian master equation
\begin{align}\label{eq:master}
\dot{\rho} = -i [H,\rho] + \hat\Gamma \rho 
\equiv \hat{\mathcal{L}} \rho.
\end{align}
In this expression, $\hat\Gamma \cdot$ is the superoperator for the different possible dissipation channels acting on the system and which can be expressed in standard Lindblad form as~\cite{gardiner:2004b}
\begin{align}
\hat\Gamma \rho = \sum_j \gamma_j \hat{ \mathcal{D}}[a_j]\rho,
\label{eq:disspation}
\end{align}
with $\hat{\mathcal{D}}[a_j]\rho = a_j \rho a_j\dag - \{a_j\dag a_j, \rho \}/2$ and $\gamma_j$ the damping rate for channel $j$ associated to the system operator $a_j$. 

The formal solution to this equation can be expressed as the time-ordered exponential
\begin{align}
	\rho(t) = 
	\mathcal{T}\! \exp 
	\left\{  
		\int_{0}^{t} \mathrm{d} t'\, \hat{\mathcal{L}}(t') 
	\right\} \rho(0).
\end{align}
Taking advantage of the piecewise constant nature of the controls, this can be written more simply as 
\begin{align}
\rho(T) = \hat{L}_N \ldots \hat{L}_j \ldots \hat{L}_1 \rho(0),
\end{align}
with the evolution superoperator defined from time $(j-1)\Delta t$ to time $j\Delta t$ as,
\begin{align}
\hat{L}_j\cdot = \exp \left\{ -i \Delta t \,(\,[H_j\, , (\cdot) ] + i\hat\Gamma\cdot\,)\, \right\}
\end{align}
where $H_j = H_0 + \sum_k u_k( j ) H_k$ is the time-independent Hamiltonian associated to the $j^\mathrm{th}$ time step.

For many control problems, the performance index can be expressed as a function of operator averages or, alternatively, as the overlap between a final state $\rho(T)$ and a target state. In both cases, the resulting performance index takes the form
\begin{align}
\Phi = \text{Tr} \Big( \sigma \hat{L}_N \ldots \hat{L}_1 \rho(0) \Big), \label{eq:phi}
\end{align}
where $\sigma$ is either the target state or an operator whose expectation value is evaluated. In the former case, if the target state is pure this figure of merit is bounded between 0 and 1, with $\Phi = 1$ for $\rho(T) = \sigma$. 

Taking advantage of the piecewise constant character of the evolution, the derivative of the performance index takes the form~\cite{Khaneja2005296}
\begin{align}
	\frac{\partial \Phi }{\partial u_k(j)} 
	&= 
	\mathrm{Tr} \left\{ 
	 \lambda_j(\sigma)
	 \frac{\partial \hat{L}_j }{\partial u_k(j)} 
	\rho_{j-1} \right\},
\end{align}
where 
\begin{align}
\rho_j = \hat{L}_j \ldots \hat{L}_1 \rho(0)
\label{eq:rho_j}
\end{align}
is a forward-in-time evolved density matrix, while 
\begin{align}\label{eq:lambda}
\lambda_j(\sigma) = \hat{L}_{j+1}\dag \ldots \hat{L}_N\dag \sigma
\end{align}
is the backward-in-time evolution from the final target state. To first order in $\Delta t$ the derivative of the $j^\mathrm{th}$ time-evolution operator is~\cite{schulte11opengrape}
\begin{equation}
	\frac{\partial \hat{L}_j\cdot }{\partial u_k(j)} 
	\approx 
	-i \Delta t \,   [ H_k, \, (\hat{L}_j\cdot) ] . \label{eq:partialuk}
\end{equation}
Approximation of the gradient to higher-order in $\Delta t$ can improve convergence of the optimization~\cite{Fouquieres:2011fk}. 
Moreover, for simplicity, we have considered the controls to be parameters of the Hamiltonian only. This approach can, however, be adapted to allow for control over the dissipation rates $\gamma_j$~\cite{PhysRevA.90.052331}.

Finally, the derivative of the performance index is 
\begin{equation}
	\frac{\partial \Phi }{\partial u_k(j)} =
-i \Delta t \,
\mathrm{Tr} \big\{ 
 \lambda_j(\sigma)
    [ H_k, \, \rho_j ]
 \big\}. \label{eq:dphidu}
\end{equation}
Thus, evaluating the gradient of the performance index requires the calculation of the forward-in-time evolved states $\rho_j$ and of the backward-in-time evolved targets $\lambda_j(\sigma)$. The analytical result of Eq.~\eqref{eq:dphidu} is the core of the GRAPE algorithm~\cite{Khaneja2005296}. The standard approach to obtain these states, $\rho_j$ and $\lambda_j(\sigma)$, is to express the density matrices and the master equation in Liouville space~\cite{schulte11opengrape}. For a system with Hilbert space dimension $d$, the superoperators then take the form of $d^2 \times d^2$ matrices and the $N$ evolution operators $\hat L_j$ are obtained by computing matrix exponentials of these matrices. While simple to implement, this procedure is numerically intensive for moderate to large system sizes.

\section{Open GRAPE with Runge-Kutta integration}
\label{sec:implementation}

Rather than relying on direct matrix exponentiation, we present here an approach based on numerical integration of the master equation using a standard Runge-Kutta routine. This approach is not a unique method to avoid the matrix exponentiation for optimal control~\cite{goerz2015optimizing, gutknecht2007brief, tal2007restart}, but below we argue that, even for moderate Hilbert space dimension, $d$, this simple Runge-Kutta routine leads to useful computational speedups compared to performing matrix exponentials as used in standard implementations of open GRAPE~\cite{schulte11opengrape}.

With this method, the forward-in-time propagation is performed by numerical integration of the differential equation
\begin{align} \label{eq:rkmaster}
d\rho = \hat{\mathcal{L}}\rho \, dt 
\end{align}
starting from the initial state $\rho(0)$ using standard Runge-Kutta routines. In practice, the integration is done in a stepwise manner to obtain $\rho_j$ for all values of $j$. In other words, Eq.~\eqref{eq:rkmaster} is integrated for a time $\Delta t$ from the initial state $\rho_0$ to obtain $\rho_1$, which is saved for later use. Then $\rho_1$ is used as initial state and integrated for a time $\Delta t$ to obtain $\rho_2$, and so on. Similarly, the backward-in-time propagation is performed by numerical integration of the master equation 
\begin{align}
-d\lambda = \hat{\mathcal{L}}^\dagger\lambda \, (-dt) ,
\end{align}
which is also solved stepwise but backward-in-time, such that $\lambda(t-\delta t) =  \lambda(t) + \hat{\mathcal{L}}^\dagger \lambda(t)(- \delta t)$ with $\delta t$ as a small numerical step, from the initial (target) state $\lambda_N = \lambda(T) = \sigma$.  Backward-in-time integration for a time $\Delta t$ leads to $\lambda_{N-1}$ which is then used as the next initial state and, continuing this way, all $\lambda_j$ are obtained. With $\rho_j$ and $\lambda_j$ calculated, the derivative given in Eq.~\eqref{eq:dphidu} is readily evaluated using the saved $\rho_j$ and $\lambda_j$.

\subsection{Complexity analysis}
\label{sec:complex}

We now turn to an evaluation of the scaling with system size $d$ of the standard approach versus the present Runge-Kutta integration method. For simplicity, we neglect the efficiency gain that can be obtained in both cases from taking advantage of the sparse character of matrices. We also take the complexity of the multiplication and exponentiation of $n\times n$ matrices to be $\mathcal{O}(n^3)$. Better scaling can be obtained from state-of-the-art algorithms, resulting in improvements for both the standard approach and the present Runge-Kutta integration method.

In the standard approach were the density operator is represented as a vector, the matrix exponentiation involved in computing the superoperators $\hat L_j$ of dimensions $d^2 \times d^2$ has a complexity $\mathcal{O}(d^6)$. For the ${N}$   piecewise constant steps of the controls, the total  complexity is therefore
\begin{equation}
	\mathcal{C}_{\mathrm{exp}} = \mathcal{O}\left( {N} \times d^6  \right).
	\label{eq:Cexp} 
\end{equation}
In contrast, the Runge-Kutta integration approach described here requires the products of operators represented by $d \times d$ matrices. One caveat of this method is that the calculation is specific to the given input state $\rho(0)$. The complexity of this approach can then be estimated as
\begin{equation}
	\mathcal{C}_{\mathrm{RK}} = \mathcal{O}\left( n_s \, n_{\mathrm{RK}} \times  d^3 \right),
	\label{eq:RK} 
\end{equation}
where $n_s$ is the number of input states to be considered and $\nRK$ the number of Runge-Kutta steps during the full time evolution. Improvement over the standard Liouville space approach is thus expected for system size \mbox{$d  \gg (n_s \nRK/N)^{1/3}$}. Importantly, the numbers  $n_s$,  $\nRK$ and $N$ are often independent of system size, suggesting a computational speedup for  large Hilbert spaces. When considering bandwidth filtered controls, where the $N$ controls are replaced by $M\gg N$ sub-pixels in order to approximate a smooth function~\cite{PhysRevA.84.022307} (see Appendix~\ref{sec:gaussian}), computational speedup is expected for even smaller Hilbert space sizes.  
Note that Eq.~\eqref{eq:RK} assumes that the Runge-Kutta integration is performed using matrix multiplications with complexity $\mathcal{O}(d^3)$. As mentioned previously, there is an alternate representation of the master equation Eq.~\eqref{eq:master} where $\rho$ is a vector of dimension $d^2 \times 1$ and the Lindbladian is a $d^2 \times d^2$ matrix. In that case, the Runge-Kutta integration requires matrix-vector multiplication of complexity $\mathcal{O}(d^4)$ reducing the speedup.

A second advantage of the present approach, not captured by this simple  analysis, is the reduced memory usage since superoperators in Liouville space of matrix size $d^2\times d^2$ are never created nor stored in memory. This reduced memory requirements by the Runge-Kutta approach is independent of the representation of the density matrix used for the integration.

The optimization of an arbitrary process requires averaging the performance index over $n_s = d^2$ input states spanning the full Liouville space~\cite{goerz14open}. 
However, this is not the case for many processes where we can expect $n_s \ll d^2$. 
In particular, average over only three appropriately chosen input states is required to optimize a unitary process in the presence of dissipation~\cite{goerz14open}. 
Another related issue is that for a general process there may be up to $d^2$ dissipators in Eq.~\eqref{eq:disspation}, which would result in the scaling of $\mathcal{O}(n_s \, n_{\mathrm{RK}} \times  d^5 )$. 
However for many problems of practical interest, such as the one presented in the next section, only a few dissipators are needed.

Estimating $\nRK$ is a more difficult task since, with adaptive integration step size, the number of integration steps is parameter and problem dependent~\cite{Galassi:2011uq}. As an example, for the reset process described in Sec.~\ref{sec:clear}, we observe that $\nRK/M \sim 10-100$ depending on the chosen value of $M$. Given that $n_s = 2$ for the reset problem, we therefore expect significant speedup even for moderate Hilbert space size of $d\sim 10$. 

Finally, we note that the Runge-Kutta approach presented here is only efficient if we perform a GRAPE-type concurrent update of the controls. In the case of a Krotov-type update where only one control is updated at each step of the optimization algorithm~\cite{PhysRevA.84.022305}, the complexity of the present method and of the approach where the density matrix is represented as a vector are expected to be similar. Indeed, the latter approach allows to reuse most of the calculated exponentials between updates. Consequently, here we consider a GRAPE-type update where all controls are updated concurrently.

\begin{figure}[tb]
	\includegraphics[width=0.9\linewidth]{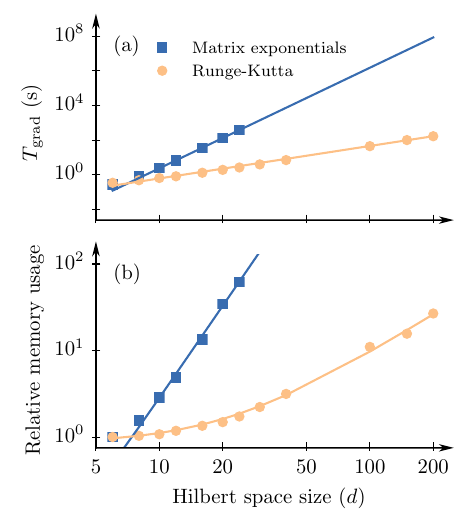}
	\caption{
	(a) Average runtime per evaluation of the performance index and its gradient ($T_{\mathrm{grad}}$). Points are numerical data obtained from the averaging over 100 optimization iterations. In the matrix exponentiation case, a first-order GRAPE implementation based on the control module of QuTiP is used~\cite{Qutip2}.
	(b) Average memory usage (RAM) during a GRAPE iteration normalized by the $d=6$ value. Points are numerical data obtained using the ``memory\_profiler'' package (version 0.47).
	For both panels, solid lines are power law fits. A constant background due to program overhead is included in the fits of the second panel (see text).
	The matrix-exponential calculations were only done for a Hilbert space size $d\leq 24$ due to limitations of RAM.
	%
	From an initial state $\ket{\alpha} \otimes \ket{e}$, with the cavity in a coherent state with $\alpha = \sqrt{d/8}$ and the qubit in the excited state,
	 we use the two quadratures of the cavity drive to perform an active reset (target state $\ket{0} \otimes \ket{e}$) in a time $T =\pi/g$ separated in $N=200$ steps.
	The qubit and cavity are coupled through a standard Jaynes-Cummings interaction with coupling strength $g = 100 \kappa$. Additional parameters are $\omega_d=\omega_r$ and $\Delta = 10 \kappa$ (see Sec.~\ref{sec:clear} for definitions).
	%
	}
	\label{fig:performance}
\end{figure}

\subsection{Performance of implementation}

As a verification of the above algorithmic complexity arguments, we consider the performance of our Runge-Kutta-based open GRAPE implementation for a simplified version of the resonator reset control problem considered in the following section.
More precisely, we consider a cavity-qubit system coupled through a standard Jaynes-Cummings interaction. Starting from an initial state where the cavity is in a coherent state and the qubit in the excited state, we optimize a drive in order to empty the cavity in a time $T$, without perturbing the qubit state, such that the target final state is $\ket{0}\otimes\ket{e}$.
This problem corresponds to a conditional reset where the qubit state is fixed leading to a single initial and final state, i.e. $n_s =1$ in Eq.~\eqref{eq:RK}. In the following section, the case of unconditional reset, where the reset protocol is independent of the qubit state, will be considered.

Fig.~\ref{fig:performance}(a) presents $T_{\mathrm{grad}}$, the average runtime for the evaluation of $\Phi$ and its gradient as a function of Hilbert space size $d$ for both our Runge-Kutta-based implementation of open GRAPE (orange circles) and a standard matrix exponentiation-based implementation (blue squares)~\cite{Qutip2}. 
In both cases, the gradient is calculated using a first order approximation as in Eq.~\eqref{eq:partialuk}.
As the approaches considered are implemented using different programming languages leading to different runtime overheads, the relevant quantity in this figure is the scaling of runtime with respect to the Hilbert space size $d$ rather than the absolute times.
The solid lines are power law fits to the numerical data, $T_{\mathrm{grad}}\propto d^\xi$ with exponent $\xi$. 
Exponents obtained are in close agreement to the previous analysis, with $\xi=5.8$ for the matrix exponentiation case, and $\xi=1.9$ for the Runge-Kutta-based approach. Note that this significant polynomial speedup is better than expected from the analysis in Sec.~\ref{sec:complex} due to the use of sparse matrix properties in our implementation.

Fig.~\ref{fig:performance}(b) presents the average memory (RAM) used during a GRAPE iteration relative to the memory usage of the $d=6$ case for each implementation. Performing again power law fits, but allowing for a constant background to take into account possible memory overheads, one finds the exponents $\xi=3.5$ for matrix exponentiation and $\xi=1.5$ for our Runge-Kutta implementation. Hence, as expected from the above complexity analysis, the memory requirement of the matrix exponentiation is much greater than the Runge-Kutta approach, as it requires the storage of propagators as $d^2 \times d^2$ matrices, limiting considerably the Hilbert space sizes on a standard computer.


\section{Application to resonator reset}
\label{sec:reset}

As an application of this open GRAPE implementation, we consider the problem of active reset following qubit readout in circuit QED \cite{wallraff2004strong, PhysRevA.69.062320}. Before presenting numerical results, we first briefly review qubit readout in this system and present the active reset problem.

\subsection{Readout and reset in circuit QED}\label{sec:clear}

Circuit QED is characterized by the strong electric-dipole coupling $g$ between a superconducting qubit of frequency $\wa$ and a microwave resonator of frequency $\wc$. In the dispersive regime, where the qubit-resonator detuning $|\Delta| = |\wa-\wc| \gg g$, the system is described by the effective Hamiltonian ($\hbar$ = 1)~\cite{PhysRevA.69.062320}
\begin{align} \label{eq:Hdrive}
H_0 =  ( \wc + \chi \sigma_z) a\dag a + \frac{\wa}{2}\sigma_z + \varepsilon(t) \left[ a\dag e^{-i \wdr t} + \mathrm{h.c.} \right], 
\end{align}
where $\chi = g^2/\Delta$ is the dispersive shift and $\mathrm{h.c.}$ stands for hermitian conjugate. The last term represents a drive on the cavity of amplitude $\varepsilon(t)$ and frequency $\wdr$. Because of the dispersive coupling, the cavity frequency is shifted by $\pm \chi$ depending on the state of the qubit. Under drive, the time-evolution leads to a qubit-state dependent population and/or phase of the cavity state. 
This dependency can be resolved by homodyne detection of the photons leaking out of the cavity at a rate $\kappa$, leading to a qubit measurement.

In order to include cavity damping in our calculations, we use the master equation
\begin{equation}
	\dot \rho = -i \left[H \,,\, \rho\right] + \kappa \hat{\mathcal{D}}\left[a \right]\rho,
	\label{eq:ME}
\end{equation}
where $\kappa$ is the cavity decay rate associated to the dissipator $\hat{\mathcal{D}}[a]\rho = a \rho a\dag - \{a\dag a, \rho \}/2$. Under a constant drive of amplitude $\varepsilon$, the steady-state solution in the dispersive regime (i.e. $H = H_0$) of this master equation leads to the qubit-state dependent intracavity average photon number
\begin{equation}
\bar n_{g/e} = \frac{ \varepsilon ^2}{(\wc \pm \chi - \wdr)^2 + (\kappa/2)^2}.
\end{equation}

Here, we are concerned with the return to vacuum state once the measurement is completed. 
The common approach of passive reset is to wait for a time $T \gg 1/\kappa$ for the photons to naturally escape from the resonator. We use our implementation of open GRAPE to find an optimal $\varepsilon(t)$ to speed-up this process to times smaller than $1/\kappa$ through an active process.

When driving at a frequency $\wc \approx \wdr$, the average number of photons is independent of the qubit state and an active reset is easily obtained by changing the phase of the drive. However, active reset is not as simple when considering the nonlinear corrections to the dispersive Hamiltonian. The first of these corrections is a qubit-induced nonlinearity of the cavity described by the Hamiltonian\footnote{In the two-level approximation of circuit QED, the sign of this nonlinear corrections is qubit-state dependent, with $H_K \propto \sigma_z$. However, in the more complete multilevel treatment, the Kerr nonlinearities $K_{g}$ ($K_e$) of the resonator for a qubit in the ground (excited) state can have the same sign~\cite{PhysRevLett.105.100504}. In particular, for the parameters considered here the Kerr nonlinearities have the same sign and are of similar amplitudes for both qubit states~\cite{mcclure2015rapid}. For simplicity, we consider $K_e \approx K_g$. }~\cite{PhysRevLett.105.100504, bourassa:2012a, nigg:2012a}
\begin{equation}
H_K = K (a\dag a)^2, \label{eq:kerr}
\end{equation}
with $K$ the Kerr-nonlinearity, which is negative in superconducting quantum circuits. This correction makes exact analytical solutions of the active reset problem difficult as it leads to nonlinear equations of motions for the resonator state. This nonlinearity can moreover lead to vastly different qubit-state dependent resonator states, something that has been exploited for qubit readout, e.g., in the Josephson bifurcation amplifier~\cite{JBA_review}. Here, because of this nonlinearity, a reset pulse more complicated than in the purely dispersive case is found to be necessary~\cite{mcclure2015rapid}.

In the next section, we present numerical results for active cavity reset based on the experimental parameters reported in Ref.~\cite{mcclure2015rapid}. For these calculations, we use the master equation of Eq.~\eqref{eq:ME} with Hamiltonian $H = H_0 + H_K$ and the experimentally relevant parameters $\chi = 2\pi \times 1.3$~MHz, $K = -2\pi \times 2.1$~kHz and $\kappa = 2\pi \times 1.1$~MHz, corresponding to a photon decay time of $T_\kappa = 1/\kappa = 145$~ns. Moreover, to help in making comparisons, we will express the drive strength in similar terms as in Ref.~\cite{mcclure2015rapid} by introducing the normalized drive power $P_\mathrm{norm} = P/P_\mathrm{1ph}$, where $P$ is the applied drive power and $P_\mathrm{1ph}$ is the drive power leading to an average steady-state resonator population of one photon. With the above parameters, we numerically identify the corresponding driving amplitude $\sqrt{P_\mathrm{1ph}} = 2\pi \times 1.595$~MHz such that $\varepsilon = \sqrt{P_{\mathrm{norm}}P_\mathrm{1ph}}$.

\begin{figure}[t]
\includegraphics[width=0.9\linewidth]{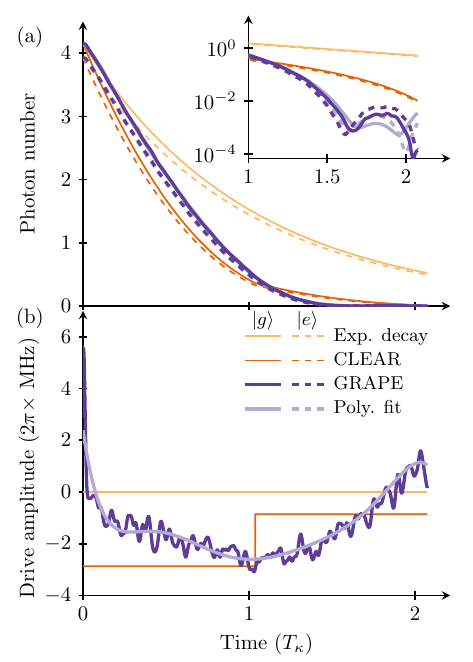}
\caption{
(Color online) (a) Average photon number during resonator reset procedures following a readout process with drive power $P_{\mathrm{norm}} = 4$. 
The solid (dashed) lines indicate results for the qubit in the ground (excited) state. See panel~(b) for legend. The inset shows the average photon number for $T_\kappa >1$ on a logarithmic scale to allow for better comparison of the reset schemes. (b) Pulse shapes for the resonator reset procedures used in panel~(a). System parameters and the Hamiltonian are described in Sec.~\ref{sec:clear}. Additional parameters for the GRAPE algorithm includes a control duration $\Delta t = 1$~ns and Gaussian filtering with bandwidth $\omega_B/2\pi = 100$~MHz and subpixel duration $\delta t = 0.1$~ns (see Appendix~\ref{sec:gaussian} for parameter definitions).}
\label{fig:expcleargrape}
\end{figure}

\subsection{Active reset using open GRAPE}
\label{sec:fasterreset}
We now turn to a numerical study of active resonator reset using the open GRAPE implementation introduced in Sec.~\ref{sec:implementation}. For simplicity, we assume the measurement preceding the resonator reset to be quantum non-demolition and, thus, consider the qubit's state to be fixed throughout the process. As a result, we can replace the operator $\sigma_z$ by the number $\pm 1$ in Eq.~\eqref{eq:Hdrive}.

As we seek an active reset protocol independent of measurement outcomes, the performance index used for the open GRAPE optimization is averaged over the two qubit states. Following Eq.~\eqref{eq:phi}, the  simplest performance index is 
\begin{align}\label{eq:Phi0}
\Phi = \sum_{i=g,e} \text{Tr}\left\{\rho_T \rho_{i}(T) \right\}, 
\end{align}
with $\rho_{i=g,e}(t)$ the qubit-dependent resonator state and $\rho_T = \ket{0}\bra{0}$ the target (vacuum) state. Here, $\rho_{i=g,e}(t=0)$ are the qubit-dependent resonator states following a measurement pulse $\varepsilon(t)$ of duration $T_m$ similar to that used in Ref.~\cite{mcclure2015rapid}. In our simulations, the resonator is initialized to the vacuum state at time $t = -T_m$, the state is then time evolved using the master equation, Eq.~\eqref{eq:ME} with $H = H_0 + H_K$, leading to the qubit-dependent states, $\rho_{i=g,e}(t=0)$. Starting from these states and using the same master equation, the open GRAPE algorithm is then used to optimize the unconditional reset pulse shape $\varepsilon(t)$ for $t \in (0,T)$, with $\varepsilon(t=0)$ fixed by the measurement pulse and $\varepsilon(t=T)=0$.

Using the parameters of the previous section, Fig.~\ref{fig:expcleargrape}(a) compares the average intracavity photon number as a function of time under various resonator reset schemes. 
In particular, the passive reset (thin light orange curves) is compared to GRAPE optimized active reset (thick dark purple curves) for duration $T = 300~\mathrm{ns} \approx 2 T_\kappa$. 
While there is still significant resonator population after a wait time $T\gtrsim 2 T_\kappa$ in the passive case, the GRAPE optimized pulse empties the cavity independently of the qubit state. More precisely, the log-scale inset, shows that the optimized pulse shape brings the photon number below $10^{-4}$ while in the same time passive reset leads to a residual average photon population close to 1. The numerically found pulse shape corresponding to these results is the thick dark purple line in Fig.~\ref{fig:expcleargrape}(b). It shows a fast oscillating behavior on top of a slowly evolving envelope. Importantly, the quality of the reset is only marginally affected by these rapid oscillations. Indeed, as shown by the thick light purple lines in both panels, a polynomial fit to the optimized pulse shape essentially leads to changes in the average photon number that are only visible on the logscale inset of Fig.~\ref{fig:expcleargrape}(a). This indicates that a complex pulse shape is not essential to obtain good performance, and that the solution may be amenable to regularization, whereby penalties are added to the objective function (for instance to penalize rapid changes in time) in order to make the result simpler and/or more robust. As pointed out in Ref.~\cite{PhysRevLett.113.010502}, the effect of noise in the control pulse, similar to fast oscillations observed here, is negligible if the noise level is below the error in reaching the target state.

As a comparison, the thin dark orange lines in Fig.~\ref{fig:expcleargrape} correspond to the average photon number and pulse shape used in an optimized two-steps active reset similar to the so-called CLEAR pulse introduced in Ref.~\cite{mcclure2015rapid}. Compared to CLEAR, the GRAPE pulse shape leads to a smaller residual photon population of the cavity in $T = 300$~ns $\sim 2 T_\kappa$. Importantly, because photon decay under GRAPE optimized pulse shapes is far from exponential, in the example of Fig.~\ref{fig:expcleargrape} the cavity is already close to having reached its final state at a time $\sim 220$~ns. This suggests that faster resets are possible.


\begin{figure}[tb]
\includegraphics[width=0.9\linewidth]{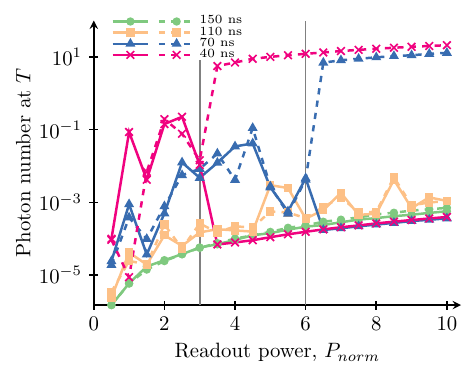}
\caption{
Average photon number at the end of the active reset pulse as a function of the readout power $P_{norm}$. The results are shown for pulses of duration {$T = $} 150~ns, 110~ns, 70~ns and 40~ns, corresponding to $T \approx 1.04~T_k, 0.76~T_\kappa, 0.48~T_\kappa, 0.28~T_\kappa$. The solid (dashed) line is the average final photon number for the qubit in the ground (excited) state. The vertical gray lines indicate the failure points for the 70~ns and the 40~ns optimizations.} 
\label{fig:finaln}
\end{figure}

To further speed-up the process, we follow the insight from DRAG and optimize over two quadratures of the drive~\cite{PhysRevLett.103.110501}. In a frame rotating at the drive frequency, the last term of the dispersive Hamiltonian of Eq.~\eqref{eq:Hdrive} is then replaced by
\begin{align}
H_{d} = \varepsilon_X(t) (a\dag + a) + i\varepsilon_Y(t) (a\dag - a).
\end{align}
Results from optimization of these two quadrature are presented in Fig.~\ref{fig:finaln}, which shows the average photon number at the final pulse time $T$ for increasing measurement power $P_\mathrm{norm}$. As an initial guess, the $X$ quadrature is set to the CLEAR pulse shape and the $Y$ quadrature is randomly set. 
These results are shown for four different values of $T$, ranging from $150~\mathrm{ns}\sim1.04 T_\kappa$ (green circles) to times as short as $40~\mathrm{ns}\sim0.3 T_\kappa$ (red $\times$). 
Following the convention of Fig.~\ref{fig:expcleargrape}(a), the full lines correspond to the qubit ground state and the dashed lines the qubit excited state. Unsurprisingly, the general trend is an increase of the residual photon number with $P_\mathrm{norm}$. However, for $150~\mathrm{ns}\sim1.04~T_\kappa$, the optimization results in residual population as small as $10^{-3}$ at high power $P_\mathrm{norm} = 10$. 

The difficulty of the open GRAPE algorithm to converge with decreasing $T$ is made apparent with the large fluctuations of the residual photon number with $P_\mathrm{norm}$. Despite this, and quite remarkably, final populations of less than $10^{-3}$ photons are obtained for reset times under $T_\kappa$ and all $P_\mathrm{norm}$ values considered. The complexity in converging becomes more apparent at very short times where we observe large fluctuations and large separations between the results obtained for the two qubit states. These branchings, corresponding to a change in the optimization landscape as a function of $T$ and $P_{\mathrm{norm}}$ \cite{sorensen2015exploring}, are illustrated by vertical gray lines for the two shortest values of $T$. Beyond the branching time, the optimization only finds a good solution for the qubit in the ground state. This is a result of the sign of the Kerr-nonlinearity $K$. Indeed, the effective detuning from the drive at high drive power is smaller when the qubit is in the ground state, thus, changing the sign of $K$ leads to finding low photon number solutions when the qubit is rather in the excited state.

\begin{figure}[tb]
\includegraphics[width=0.9\linewidth]{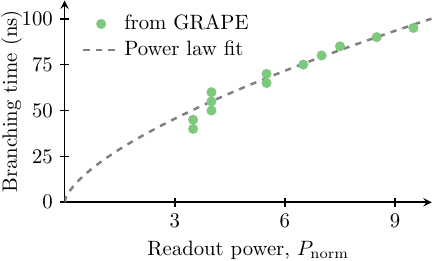}
\caption{
The green dots are the numerical speed limit extracted from the open GRAPE optimizations. We define here the speed limit as the time where the optimization fails, corresponding to the branching points indicated by gray lines in Fig \ref{fig:finaln} for the 40~ns and 70~ns curves. The dashed gray line is a power law fit,~$\propto (P_{\mathrm{norm}})^\xi$, to the data with $\xi  = 0.65$.
} \label{fig:speed}
\end{figure}
Fig.~\ref{fig:speed} presents this branching time as a function of $P_{\mathrm{norm}}$. As illustrated by the dashed line, this failure time follows a simple power law behavior. This is reminiscent of a quantum speed limit, which here corresponds to the minimal time $T$ in which the optimization can be successful~\cite{PhysRevLett.103.240501, goerz11speed}. For pure state evolution, the quantum speed limit can be expressed analytically in terms of the mean value and the variance of the energy~\cite{PhysRevA.67.052109, PhysRevLett.103.160502}. Expressions have also been obtained for open processes~\cite{PhysRevLett.110.050403, PhysRevLett.110.050402}. The observed simple behavior with $P_{\mathrm{norm}}$ suggests that analytical expressions could also be obtained for the reset problem. We note, however, that variations in the initial guess for the controls, cost function or optimization algorithm could lead to faster reset times~\cite{sorensen2015exploring}, and that the results of Fig.~\ref{fig:speed} therefore do not represent an absolute speed limit.

\begin{figure}[b]
\includegraphics[width=0.9\linewidth]{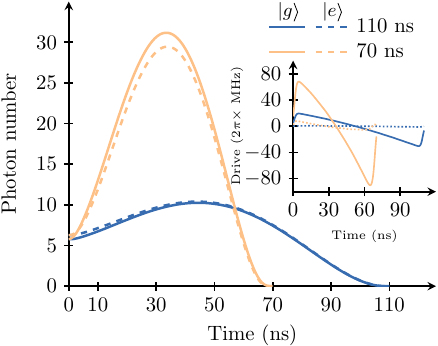}
\caption{Photon number as a function of time during the active reset pulse for a pulse duration of 110~ns and 70~ns. The solid (dashed) line is for the qubit in the ground (excited) state. The inset shows the corresponding Gaussian filtered drives. The solid lines of the inset is the $X$-drive, while the dotted is the $Y$-drive.}
\label{fig:timeevo}
\end{figure}
\begin{figure}[tb]
\includegraphics[width=0.9\linewidth]{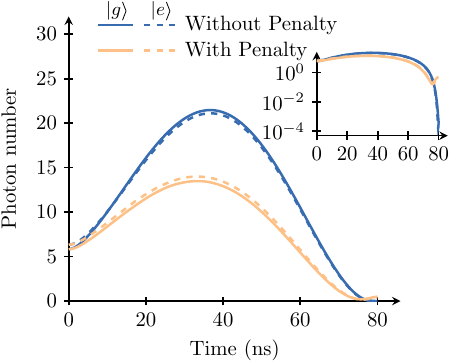}
\caption{Photon number as a function of time for optimized drives with (light orange curves) and without (dark blue curves) the photon number penalty, $\Phi_p$, included. The solid (dashed) curves are for the qubit in the ground (excited) state. The inset displays the same data with a logarithm photon number axis. The parameters are the same as Fig.~\ref{fig:expcleargrape}. We use the penalty weight $\beta = 0.2/T$.}
\label{fig:photonapp}
\end{figure}
To gain more insights on the optimization, Fig.~\ref{fig:timeevo} presents the average photon number as a function of time and the corresponding pulse shapes obtained from GRAPE (inset). These results are shown for $T= 70$~ns (light orange lines) and $T=110$~ns (dark blue lines) with a readout power of $P_\mathrm{norm} = 6$. Both pulse shapes are similar and are reminiscent of a smoothed CLEAR pulse~\cite{mcclure2015rapid}. The $Y$ quadrature also appears to have minimal impact and is always close to zero. For both of the final times $T$, the average photon number first increases from its initial value of $\sim 6$ before decreasing to the value shown in Fig.~\ref{fig:finaln}. This increase is particularly notable for the short pulse time $T=70$~ns and points to the difficulty in converging as the reset time $T$ is decreased. In practice, this large photon population can lead to a breakdown of the dispersive approximation used here and to a departure from the quantum non-demolition character of the dispersive readout~\cite{PhysRevA.79.013819}. With the parameters used here, this breakdown is expected to occur for $T=70$~ns where the average photon number exceeds the critical photon number $n_\mathrm{crit} = (\Delta/2g)^2 \sim 29$ for a short period of time.

To prevent this large photon number increase, a penalty $\Phi_p$ related to the intracavity photon number can be added to the performance index such that
\begin{align}\label{eq:PerformancePenalty}
\Phi = \Phi_0 - \beta \Phi_p,
\end{align}
with $\Phi_0$ defined by Eq.~\eqref{eq:Phi0} and $\beta$, a constant weighting the penalty, which is determined by trial and error. To penalize large photon populations we take
\begin{align}
\Phi_p = \sum_{i=g,e} \int_0^T \text{Tr} \left\{a\dag a \,\rho_i(t) \right\} dt. \label{eq:phi_p}
\end{align}
Details about the numerical implementation of $\Phi_p$ and its derivative with respect to the controls can be found in Appendix~\ref{app:photonpenalty}. Results for optimization with this modification of the performance index are presented in Fig.~\ref{fig:photonapp} for $T= 80$~ns and $P_\mathrm{norm} = 6$. For these values, the optimization without penalty reaches a final photon population of $10^{-4}$ but reaches close to 25 photons in the transient dynamics. On the other hand, using Eq.~\eqref{eq:PerformancePenalty}  with the initial value of the pulse given by the results obtained without penalty, the transient photon number can be kept well below $n_\mathrm{crit}$. This is however achieved at the cost of an increase of the final photon number to $\sim10^{-1}$. These results for the photon penalty may be improved by considering more diverse initial pulse shapes probing a larger region of the optimization space. In addition, a more systematic study of the role and optimal value of the weight $\beta$ could improve the results.

\section{Conclusion and outlook}
\label{sec:conclusion}
 
We have shown an implementation of the GRAPE algorithm for open quantum systems that circumvents the usual explicit calculation of matrix exponentials.This implementation is advantageous when optimizing quantum processes in large open quantum systems leading to reduced computation times and memory requirements compared to standard implementations of open GRAPE based on matrix exponentials.

As an example of this approach, we have demonstrated an optimized reset protocol for a readout resonator in circuit QED. As the reset time limits the repetition time of current experiments, rapid qubit reset after readout is of high practical importance. Moreover, rapid qubit recycling can  be advantageous in the implementation of quantum algorithms~\cite{martin2012experimental}. Furthermore the results of our optimization may be directly applied to protocols that rely on repetitive qubit readout in circuit QED, e.g., in quantum feedback schemes \cite{riste2013deterministic, andersen2015closing, PhysRevLett.112.170501} or in quantum error correction protocols \cite{PhysRevA.86.032324,corcoles2015demonstration}. The numerical optimization presented in this work presents a reset scheme that significantly reduces reset time compared to passive reset. Moreover, this study pinpoints the issues occurring when extremely short reset times are sought and yields a branching point beyond which the optimization algorithm fails to find a qubit-state-independent solution. We find that this branching follows a power law as a function of the readout power, indicating a 
relation between the system's energy and the shortest time required to achieve the target states. Finally, we identify that our scheme can be readily extended to include additional constraints such as a penalty on large average photon numbers in order to keep the cavity population below the critical photon number set by the dispersive approximation. 

While resonator reset in the dispersive regime of circuit QED serves as an instructive study, we emphasize that  this implementation of GRAPE may have much broader use. As a second practical example, our approach has also been recently applied by some of us to fast cat states generation in nonlinear resonators~\cite{Puri:2017ee}. Following recent experimental results, our work could be expanded to study resonator reset in the strongly nonlinear regime of circuit QED~\cite{bultink:2016a}. Our approach appears ideally suited to simulate the large Hilbert space that is needed to simulate these experiments. Another application is the optimization of qubit measurement in circuit QED~\cite{PhysRevA.90.052331, PhysRevLett.114.200501}. Finally, our implementation may prove useful in optimizing unitary gates that not only works in the qubit subspace but rely on the full Hilbert space of a resonator and multiple qubits~\cite{PhysRevA.91.032325}.

\begin{acknowledgments}
The authors acknowledge valuable feedback from F.~Motzoi.
CKA and JV thank Universit\'{e} de Sherbrooke for their hospitality. SB and AB acknowledge financial support from NSERC. CKA acknowledges financial support from the Villum Foundation Center of Excellence, QUSCOPE, and from the Danish Ministry of Higher Education and Science. JV would like to thank MITACS Globalink Program for financial assistance.
Computations were made on the supercomputer Mammouth parallele II from Universit\'{e} de Sherbrooke, managed by Calcul Qu\'{e}bec and Compute Canada. 
The operation of this supercomputer is funded by the Canada Foundation for Innovation (CFI), NanoQu\'{e}bec, RMGA and the Fonds de recherche du Qu\'{e}bec - Nature et technologies (FRQ-NT). This research was undertaken thanks in part to funding from the Canada First Research Excellence Fund.
\end{acknowledgments}

\begin{appendix}

\section{Gaussian filter}
\label{sec:gaussian}

In this Appendix, we present the Gaussian filtering procedure developed by Motzoi \textit{et. al.} in Ref.~\cite{PhysRevA.84.022307}, and mentioned in Sec.~\ref{sec:grape}.
In circuit QED, while typical electronics limits the controls $\{u_k(j)\}$ to a minimal duration $\Delta t$ of a \mbox{few ns}, the limited bandwidths of control lines and pulse generators leads to a smoothed drive which can significantly modify the dynamics from the one expected from piecewise constant drives.
The filtering approach summarized here allows to incorporate these experimental constraints in the GRAPE algorithm.
The main idea of Ref.~\cite{PhysRevA.84.022307} is to calculate the dynamics using a new smoothed pulse $s_k(t) \equiv s_k[\{u_k(j)\},t]$ which is a functional of the set of controls, while still performing the optimization on the $N$ controls $\{u_k(j)\}$. 

As the GRAPE algorithm requires a piecewise constant field, this new smoothed drive $s_k(t)$ is approximated as a piecewise constant drive, with each step a subpixel of amplitude $s_{k,n}$ and duration $\delta t \ll \Delta t$. 
The set of controls, $\{u_k(j)\}$, now translates into a set of drive amplitudes, $s_{k}(n)$, for a time  $t \in [(n-1) \delta t ; \, n\delta t [$
with  $n \in \{1, 2, \dots M\}$ and $M = T/\delta t \gg N$ the number of subpixels. The controls and the smoothed drive are related by 
\begin{align}
s_{k}(n) = \sum_{j=1}^{N} T_{k,n,j}\, u_{k}(j),
\end{align}
with $T_{k,n,j}$ a transfer function matrix which act as a filter on the controls.
The derivatives of the performance index can be calculated using the chain rule
\begin{align}
\pfrac{\Phi}{u_k(j)} = \sum_{n=1}^M \pfrac{\Phi}{s_{k}(n)} \pfrac{s_{k}(n)}{u_{k}(j)},
\end{align}
where the derivative with respect to $s_{k}(n)$ can be found using Eq. \eqref{eq:partialuk}, while $\partial s_{k}(n) / \partial u_k(j)$ comes directly from the transfer matrix.

In this paper, all numerics use transfer functions based on Gaussian filters since most experimental hardware constraints can be approximated well by such a filter~\cite{PhysRevA.84.022307}. Hardware components are typically characterized by their 3dB attenuation bandwidth, $\omega_B$. Using a filter function
\begin{align}
F(\omega) = \exp (- \omega^2 / \omega_{0}^2 ),
\end{align}
with the reference bandwidth for a given control field given by \mbox{$\omega_0 = \omega_B / (-\text{ln}(1/\sqrt{2}))^{1/2} \approx \omega_B /0.5887$}, the transfer matrix can now be calculated as~\cite{PhysRevA.84.022307}
\begin{align}
T_{k,n,j} &= \int_{-\infty}^{\infty} \hspace{-0.1cm} \frac{F(\omega){}\cos\big(\omega\frac{2(n{-}1)\delta t {-}(2j{-}1)\Delta t}{2}\big)\sin(\frac{\omega\Delta t}{2})}{\pi \omega} d\omega \nonumber \\
&= \frac{ \text{erf}\Big[\omega_{0} \frac{(n{-}1)\delta t{-}(j{-}1)\Delta t}{2}\Big] - \text{erf}\Big[\omega_{0} \frac{(n{-}1)\delta t{-}j\Delta t}{2}\Big]}{2},
\end{align}
with erf being the error function.

\section{Photon number penalty}
\label{app:photonpenalty}
In this Appendix, we detail the numerical calculation of the gradient $\partial \Phi_p / \partial s_k(j)$ of the photon number penalty to the performance index $\Phi_p$ defined in Eq.~\eqref{eq:phi_p} of Sec.~\ref{sec:fasterreset}. Using Appendix~\ref{sec:gaussian}, this can be translated into $\partial \Phi_p / \partial u_k(j)$ needed for the update rule, Eq.~\eqref{eq:update}. We show that, even though $\Phi_p$ is the result of a time integration over the full duration of the reset process, the gradient can still be calculated using a single forward and a single modified backward evolution.

In order to calculate $\Phi_p$ numerically, we approximate the continuous integral of Eq.~\eqref{eq:phi_p} by a discrete sum over the subpixels defined in Appendix~\ref{sec:gaussian},
\begin{align}
\Phi_p \approx \sum_{i=e,g} \sum_{n=0}^M \delta t \text{Tr}\big( a\dag a \hat{L}_n \ldots \hat{L}_1 \rho_i(0) \big).
\end{align}
Now, we need to find $\partial \Phi_p / \partial u_k(j)$. 

The gradient of the integration over time of the mean value of an operator $A$ is in general given by
\begin{align}
\sum_{n=0}^M \delta t \frac{\partial \exv{A}_n }{\partial s_k(j)} =& \sum_{n=0}^M \delta t\, \text{Tr} \Big( A \frac{\partial (\hat{L}_n \ldots \hat{L}_1)}{\partial s_k(j)} \rho \Big) \\
=& \sum_{n=0}^M \delta t \, \text{Tr} \Big[ A\, \hat{L}_n \ldots \hat{L}_{j+1} \Big( \frac{\partial \hat{L}_j}{\partial s_k(j)} \Big) \nonumber \\
& \times \hat{L}_{j-1} \ldots L_1 \rho(0) \Big] \Theta(n-j)
\end{align}
where we have used the Heavyside step function
\begin{align}
\Theta(n) = \begin{cases}0 & \text{ if } n< 0 \\ 1 & \text{ if } n \geq 0 \end{cases}.
\end{align}
Using the linearity of the trace, we see that
\begin{align}
\sum_{n=0}^M \delta t \frac{\partial \exv{A}_n }{\partial s_k(j)} =& \text{Tr} \Big[ \Big( \sum_{n=0}^M \delta t \Theta(n-j) A \,\hat{L}_n \ldots \hat{L}_{j+1}  \Big) \nonumber \\
& \phantom{\text{Tr} \Big[} \times \frac{\partial \hat{L}_j}{\partial s_k(j)} (\hat{L}_{j-1} \ldots L_1 \rho(0)) \Big],
\end{align}
such that the last parentheses inside the trace is the same as the forward evolution used for the calculation of $\Phi_0$, while the first parenthesis is a stepwise backward evolution starting from the operator $A$. This backward evolution is equivalent to a sum over backward evolutions starting at all time steps. For example, for $j=M-2$ the parenthesis reads $ A \hat L_M \hat L_{M-1} + A \hat L_{M-1} + A = (A \hat L_M + A)\hat L_{M-1} + A$. 
Therefore we can rewrite the gradient of the photon number penalty as
\begin{align}
\frac{\partial \Phi_p}{\partial s_k(j)} = \delta t \sum_{i=e,g} \text{Tr} \Big( \zeta_{M-j} \frac{\partial \hat{L}_j}{\partial s_k(j)} \rho_{j-1}\Big),
\end{align}
with the quantities $\zeta_{M-j} = a\dag a + \hat{L}_{j+1}^{\dagger} \zeta_{M-j+1}$ defined recursively starting from $\zeta_M = a^\dagger a$ and $\rho_{j} = \hat{L}_{j} \ldots L_1 \rho(0)$ as defined in Eq.~\eqref{eq:rho_j}. The derivative $ \partial \hat{L}_j/\partial s_k(j)$ is calculated as in Eq.~\eqref{eq:partialuk}. Thus, by adding $a^\dagger a $ to the result of the backward evolution at each timestep, the scaling of the GRAPE algorithm is not affected by this more complicated performance index and the gradient of the penalty function is obtained by the calculation of only one forward and one modified backward evolution per qubit state considered.

\end{appendix}

\bibliography{bt}

\end{document}